 \documentclass[aps,showpacs,amsmath,,preprint,preprintnumbers]{revtex4}

\usepackage{graphicx}
\usepackage{color}

\def \be  {\begin{equation}}
\def \ee  {\end{equation}}
\def \bea {\begin{eqnarray}}
\def \eea {\end{eqnarray}}

\newcommand{\nn}{\nonumber}

\begin{document}

\preprint{ECTP-2015-14}
\preprint{WLCAPP-2015-14}
\vspace*{3mm}

\title{Strangeness production in high-energy collisions and Hawking-Unruh radiation}

\author{Abdel Nasser Tawfik}
\email{a.tawfik@eng.mti.edu.eg}
\affiliation{Egyptian Center for Theoretical Physics (ECTP), Modern University for Technology and Information (MTI), 11571 Cairo, Egypt}
\affiliation{World Laboratory for Cosmology And Particle Physics (WLCAPP), Cairo, Egypt}

\author{Hayam Yassin}
\author{Eman R. Abo Elyazeed}
\affiliation{Physics Department, Faculty of Women for Arts, Science and Education, Ain Shams University, 11577 Cairo, Egypt}

\begin{abstract}

The assumption that the production of quark-antiquark pairs and their sequential string-breaking taking place through the event horizon of the color confinement determines freezeout temperature and gives a plausible interpretation of the thermal pattern of pp and AA collisions. When relating the black-hole electric charges to the baryon-chemical potentials it was found that the phenomenologically-deduced parameters from various particle ratios in the statistical thermal models agree well with the ones determined from the thermal radiation from charged black-hole. Accordingly, the resulting freezeout conditions, such as $s/T^3=7$ and $<E>/<N>=1~$GeV, are confirmed at finite chemical potentials, as well. Furthermore, the problematic of strangeness production in elementary collisions can be interpreted by thermal particle production from the Hawking-Unruh radiation. Consequently, the freezeout temperature depends on the quark masses. This leads to a deviation from full equilibrium and thus a suppression of the strangeness production in the elementary collisions. But in nucleus-nucleus collisions, an average temperature should be introduced in order to dilute the quark masses. This nearly removes the strangeness suppression. An extension to finite chemical potentials is introduced. The particle ratios of kaon-to-pion, phi-to-kaon and antilambda-to-pion are determined from Hawking-Unruh radiation and compared with the thermal calculations and the measurements in different experiments. We conclude that these particle ratios can be reproduced, at least qualitatively, as Hawking-Unruh radiation at finite chemical potential. With increasing energy, both K+/pi+ and phi/K- keep their maximum values at low SPS energies. But the further energy decrease rapidly reduces both ratios. For Lambda/pi-, there is an increase with increasing collision energy, i.e. no saturation is to be observed.

\end{abstract}

\pacs{04.70.Dy, 04.70.Dy, 14.20.Jn}
\keywords{Evaporation of black hole, Thermodynamics of black holes, Strangeness particle}

\maketitle

\section{Introduction}

In classical theory, the black holes are assumed not being able to emit particles or radiations. They are assumed to absorb, only. But the quantum mechanical effects allow black holes to create and even emit particles and radiations as if they were hot systems obeying the thermodynamical laws \cite{Hawking:1974,Hawking:1975,Unruh:1976,0704.1426}. The proposed correspondence between the hadronization processes (e.g. Ref. \cite{Tawfik:2014eba}) in the high-energy experiments and the Hawking-Unruh radiation \cite{Hawking:1974,Hawking:1975,Unruh:1976,0704.1426} from black holes, which dates back to seventieth of the last century \cite{Recami:1976} gives a plausible interpretation for the universally observed features of $e^+ e^-$, $p-\bar{p}$,  $p-p$ and  $A-A$ collisions that the particle production from these different colliding systems have characterizing thermal patterns \cite{0711.3712}.  While in $A-A$ collisions the thermalization can be understood due to the possible rescatterings between their large number of interacting partons,  in $e^+e^-$, $p-\bar{p}$, $p-p$ collisions such processes are very limited \cite{0711.3712}. The analogy of the hadronization from quantum chromodynamic (QCD) collisions and the thermal radiation from the black holes was conjectured to solve this thermalization puzzle \cite{0704.1426}.
\begin{itemize}
\item Vacuum instability through pair production which allows quantum tunnelling through the event horizon of the quarks and the gluons. This process leads to a thermal radiation at a temperature that could be determined from the quark-antiquark string-tension.
\item The partition processes as the proposed quantum tunnelling likely take place at high energy and lead to a limiting temperature similar to the one in the Hagerdorn thermal models.
\item The successive cascades and rescatterings, which are likely lost through initial state, especially in kinetic thermalization, represent a third puzzle. The stochastic QCD Hawking radiations take place in equilibrium and thus prevent the transfer of information.
\end{itemize}

The description of the hadronization process as Hawking-Unruh radiation is based on string breaking mechanism which has the correct dynamics for the particle production in  the high temperature and small chemical potential regimes. The dependence of Hawking-Unruh radiation temperature ($T_{\mathrm{HU}}$) on the chemical potential ($\mu$) and/or the angular momentum was proposed in Refs.  \cite{Chapline:1975,Recami:1976,Sivaram:1977,Grillo:1979,0704.1426,0711.3712,1409.3104,1403.3541,exact_string,Tawfik:2015fda}. The produced hadrons are assumed to be {\it born in equilibrium}, where the Rindler spacetimes are considered as a near-horizon approximation for the black-hole spacetimes \cite{exact_string}. It has been concluded that the dependence of $T_{\mathrm{HU}}$ on $\mu$ or equivalently  on the center-of-mass energy ($\sqrt{s_{\mathrm{NN}}}$) assumes that \cite{exact_string}: 
\begin{itemize}
\item the black hole mass is proportional to $\mu$ (or $\sqrt{s_{\mathrm{NN}}}$), 
\item the string tension ($\sigma$) plays the role of the running coupling constant ($\alpha_s$) \cite{Dijkgraaf:1991ba,Grumiller:2005sq}, 
\item the Hawking and Unruh temperatures are identical and 
\item the black-hole partition function likely diverges at the Hagedorn temperature \cite{Tawfik:2014eba}.
\end{itemize}

In high-energy experiments, it is conjectured that the hadrons at high temperatures can be deconfined into quark-gluon plasma (QGP). But by reducing the temperature, QGP confines back forming hadrons. At the chemical freezeout temperature ($T_{\mathrm{ch}}$), the cooling hadrons go through a state of chemical equilibration, called chemical freezing-out, in which the number of produced particles becomes fixed, i.e.  no further particles shall be produced or decay \cite{Tawfik:2014eba,Tawfik:2004vv}. It was proposed \cite{1409.3104} that the freezeout conditions, such as the entropy density normalized to $T^3$, $s/T^3$, calculated from Hawking-Unruh radiation has almost the same value as the one determined for high-energy collisions at vanishing chemical potential \cite{Tawfik:2005qn,Tawfik:2004ss,Tawfik:2005gk}
\begin{eqnarray}
\left.\frac{s}{T^3}\right|_{Q=0} = \frac{3}{8\, G^2\, M}\frac{1}{T^3_{\mathrm{BH}}(M,0)}, \label{eq:sT3bh}
\end{eqnarray}
where $M$ and $G$ being black hole mass and gravitational constant, respectively and $T_{\mathrm{BH}}(M,0)$ is the Hawking-Unruh temperature at finite $M$ and vanishing chemical potential. Assuming that the proposed analogy of the hadronization in high-energy collisions and the Hawking-Unruh radiation from black holes remains valid at finite $\mu$, we have shown this the freezeout condition \cite{Tawfik:2005qn,Tawfik:2004ss,Tawfik:2005gk} can also be obtained from Hawking-Unruh radiation from electrically-charged black-holes \cite{Tawfik:2015fda}
 \begin{eqnarray}
\left.\frac{s}{T^3}\right|_{Q \neq 0} = \left.\frac{s}{T^3}\right|_{Q=0} \left\{2 \left[1+\left(1-\frac{Q^2}{G\, M^2}\right)^{1/2}\right]^5 \left(1-\frac{Q^2}{G\, M^2}\right)^{-3/2}\right\},  \label{eq:sT3b}
 \end{eqnarray}
where $Q$ is the black-hole electric charge.

The thermal hadron production through Hawking-Unruh radiation was proposed \cite{1403.3541} as an attempt to solve the problematic of strangeness production in elementary collisions. It is seen as a mechanism determining the dependence of the freezeout temperature on the quark mass. Similar ideas have been proposed in Refs. \cite{0612151,0704.1426,0711.3712,1403.3541,Grumiller:2005sq,exact_string}. This leads to a deviation from full equilibrium and thus explains the suppression of the strangeness production in elementary collisions. In $A-A$ collisions, the average temperature diluting the quark mass effect should be introduced. Accordingly, the strangeness suppression nearly disappears. In light of this, the ratio of kaon to pion was calculated at vanishing chemical potential and found nearly equal to the maximum value measured at superproton synchrotron (SPS) energies \cite{1403.3541}. In the present work, we extend these proposals to finite chemical potentials, which are assumed being proportional to the black-hole electric charges.  In other words, the same assumption that the baryon chemical potential in high-energy collisions is correspondent to the Hawking-Unruh thermal radiation from electrically-charged black holes, which we have introduced in Ref. \cite{Tawfik:2015fda} and used in describing the $T$ - $\mu$ phase diagram \cite{Tawfik:2014eba} and confirming the freezeout conditions ($s/T^3 =7$ and $E/N \sim 1$), is assumed as an extension of the conjecture introduced in Ref. \cite{1403.3541} to explain the strangeness production in high-energy collisions at finite baryon chemical potential. From the proposed relation between $T_{\mathrm{HU}}$ and the exact-string black hole mass \cite{Dijkgraaf:1991ba,Grumiller:2005sq}, we assume a relation connecting the latter with the baryon chemical potential (heavy-ion collisions) or its analogy in electrically-charged black-holes and with the center of mass energy. For the sake of completeness, one could think of other physical properties such as spin or angular momentum ($J$) and accordingly propose their relations with mass, baryon chemical potential and temperature.

The present paper is organized as follows. The correspondence of the hadronization processes in high-energy collisions and the Hawking-Unruh radiation shall be elaborated in section \ref{sec:approach}. The extension to finite baryon chemical potential is introduced. The relation between the quark masses and the radiation temperatures at vanishing baryon chemical potential is shortly reviewed in section \ref{sec:quarkmassBH}. The relation between the exact-string black hole mass and the center-of-mass energy (chemical potential) is introduced in section \ref{sec:tempExact}. The freezeout temperatures at finite baryon chemical potential from exact-string black hole are also discussed. The energy dependence of various particle ratios determined at freezeout temperatures shall be studied in section \ref{sec:Ratios}. Section \ref{sec:res} is devoted to the results and discussion. The conclusions are elaborated in section \ref{sec:cncl}.

\section{The theoretical approach}
\label{sec:approach}

There have been various works assuming correspondence between the black-hole radiation and the hadronization process in high-energy collisions \cite{Chapline:1975,Recami:1976,Sivaram:1977,Grillo:1979,0711.3712,1409.3104,1403.3541,Tawfik:2015fda}. The mass of black hole is assumed to be related to the Unruh temperature (Rindler horizon) \cite{exact_string}. The Hawking radiation mechanism \cite{Hawking:1974,Hawking:1975} predicts that the mass of the black hole decreases with the radiation until the complete evaporation takes place \cite{gr-qc/0010055}. Such macro-canonical treatment neglects the mass-loss through the evaporation process. If the mass of the black hole becomes comparable to the radiation temperature (when the mass reaches the Planck limit), then the macro-canonical ensemble is no longer applicable (should be replaced by micro-canonical one). 

On the other hand, the high-energy experiments are designed to study the QCD under extreme conditions of high temperature and density (chemical potential), where the colliding hadrons are assumed to go through phase transitions into deconfined quarks and gluons, i.e.  quark gluon plasma (QGP). This partonic system expands rapidly and cools down drastically forming confined hadrons again, i.e.  hadronization process. With a further decrease in the temperature the hadronized system freezes out producing particles (hadrons) in the final state. The abundances of various particle species can be very well described by the statistical thermal models \cite{Tawfik:2014eba}. In a previous work \cite{Tawfik:2015fda}, we have shown that the event horizon of the color confinement of Hawking-Unruh radiation leads to universal freezeout temperature, i.e.  depending of the freezeout temperature on the chemical potential or the black-hole charge. Castorina and Satz discussed the dependence of the freezeout temperature on the quark masses at vanishing chemical potential \cite{1403.3541}. For a complete description for the black-hole radiation or equivalently the QCD hadronization, thermal and statistical conditions should be fulfilled including, for instance, the quark occupation parameters $\gamma_q$ and $\gamma_s$, for light and strange quark, respectively, and excluded volume corrections, etc. \cite{Tawfik:2014eba}. In the present work, we propose an extension to finite chemical potential (charged black-hole).

Let us assume a charged black-hole of Reissner-Nordstr\"om type \cite{Ray:1992}. The invariant space-time length element can be given as
\be
ds^2 = \left(1 - 2\frac{G M}{R_{\mathrm{RN}}} + \frac{G Q^2}{R_{\mathrm{RN}}^2}\right)\; d t^2 - \left(1- 2\frac{G M}{R_{\mathrm{RN}}} + \frac{G Q^2}{R_{\mathrm{RN}}^2}\right)^{-1}\; dr^2,
\label{RN}
\ee
where
\be
R_{\mathrm{RN}} = G\, M\, \left[1 + \left(1 - Q^2/GM^2\right)^{1/2}\right],
\label{RNradius}
\ee
and $r$ and $t$ are flat space and time coordinates, respectively. $G\simeq1/(2\sigma)= 2.5~$GeV$^{-2}$, with $\sigma=0.2~$GeV$^2$ defining the string tension. The Schwarzschild radius $R_{\mathrm{S}}=2GM$ can be reached/obtained, when the electric charge $Q$ is assumed becoming vanishing. At finite electric charge, the temperature of Hawking-Unruh radiation reads \cite{Tawfik:2015fda}
\bea
T_{\mathrm{BH}}(M,Q) &=& \frac{1}{2\pi}\, \frac{G M R_{\mathrm{RN}} - GQ^2}{R_{\mathrm{RN}} G^2 M^2} \left[\frac{1}{\left(1+\sqrt{1-Q^2/GM^2}\right)^2}\right] \nn \\ 
&=& T_{\mathrm{BH}}(M,Q=0)\; \left[\frac{4~\sqrt{1 - Q^2/ GM^2}}{\left(1 + \sqrt{1- Q^2/GM^2}\right)^{2}}\right],
\label{T-Q}
\eea
where $T_{\mathrm{BH}}(M,0)$ is the Hawking-Unruh radiation temperature from Schwarzschild black hole with $M=1.463~$GeV and $Q=0$. Details about the extension to finite baryon chemical potential can be taken from Ref. \cite{Tawfik:2015fda}. The quantities $M$, $Q$, and $\mu$ are conjectured to be related to each others \cite{0007195,0704.1426}.

From Eq. (\ref{RNradius}) and Bekenstein entropy $S=\pi R^2_{\mathrm{RN}}/G$, we obtain \cite{0704.1426}
\bea
S &=& 2 \pi G M^2 \left[1 + \left(1-\frac{Q^2}{G M^2}\right)^{1/2}\right] - \pi Q^2, \label{eq:S-RN1}\\
d S &=& \left[\frac{4 \pi}{f^{\prime}(R_{\mathrm{RN}})}\right] d M - \left[\frac{4 \pi Q}{f^{\prime}(R_{\mathrm{RN}})\, R_{\mathrm{RN}}}\right] d Q, \label{eq:dS-RN1}
\eea
where $f(R_{\mathrm{RN}})=1 - 2 G M /R_{\mathrm{RN}} + G Q^2/R_{\mathrm{RN}}^2$ and $f^{\prime}(R_{\mathrm{RN}})=2 G M /R_{\mathrm{RN}}^2 - G Q^2/R_{\mathrm{RN}}^3$ \cite{Tawfik:2015fda}. It becomes a trivial exercise to substitute $T=f^{\prime}(R_{\mathrm{RN}})/4 \pi$ in Eq. (\ref{eq:dS-RN1}) in order to get
\bea
d M - T\, d S = \left(\frac{Q}{R_{\mathrm{RN}}}\right) d Q,
\eea
which can be compared with the first law of thermodynamics,
\be
d M - T\, d S = \mu\, d Q. \label{eq:1stThrm}
\ee
Accordingly, the baryon chemical potential reads
\bea
\mu & \equiv & \frac{Q}{R_{\mathrm{RN}}} \simeq 1.4 \frac{\pi Q R_{\mathrm{RN}}}{G S}, \label{mu-Q}
\eea
where $1.4$ is an {\it ad hoc} parameter introduced to assure a good agreement with the measurements in different experiments, section \ref{sec:res}. It has been assumed  that this factor ranges between $1/G$ and $4$ \cite{0007195,0704.1426}.

Equations (\ref{T-Q}) and (\ref{mu-Q}) can be used to  map out the freezeout {\it phase-diagram} (relating temperatures to baryon chemical potentials). This is the freeze {\it phase-diagram} \cite{Tawfik:2015fda} deduced from Hawking-Unruh radiation. It can be compared to the one determined from the heavy-ion collision experiments \cite{Tawfik:2014eba}. Also, Fig. \ref{HawkingUnruhFreezeout} shall depicts all these results, as well.

For the sake of completeness, we emphasize that the physical properties of the black hole, namely $M$, the $Q$ and the spin or angular momentum ($J$), can be taken into consideration. For instance, at $J\neq0$ and $Q\neq0$ (Kerr-Newman black hole), Eq. (\ref{eq:1stThrm}) gets an extension  \cite{0704.1426},
\be
d M = T\, d S + \mu\, d Q + \Omega\, d J, \label{eq:1stThrmSpin}
\ee
where $\Omega$ is the rotational velocity and $R_{\mathrm{KN}}$ being the radius of the event horizon of this type of black hole, 
\bea
\Omega &=& \frac{4 a G}{(R^2_{KN}+j^2)}, \\
R_{\mathrm{KN}} &=& G\, M \left[1+\left(1-\frac{Q^2}{G\, M^2}-\frac{j^2}{G^2\, M^2}\right)^{1/2}\right].
\eea
with $j=J/M$ \cite{0704.1426}. The corresponding temperature is given as \cite{Li:1983,0606018}
\be
T_{\mathrm{BH}}(M,Q,J)=T_{\mathrm{BH}}(M,0,0) \left[\frac{4\left[1-(G\, Q^2+j^2)/G^2\, M^2\right]^{1/2}}{\left\{1+\left[1-(G\, Q^2+j^2)/G^2\, M^2\right]^{1/2}\right\}^2+j^2/G^2\,  M^2}\right].
\label{Kerr-Newman}
\ee
Now, we have a set of expressions for $T_{\mathrm{HU}}$ of Schwarzschild \cite{Tawfik:2015fda}, Reissner-Nordstr\"om, Eq. (\ref{T-Q}) \cite{Tawfik:2015fda}, and Kerr-Newman black holes, Eq. (\ref{Kerr-Newman}) \cite{Li:1983,0606018}.

\subsection{Quark mass and radiation temperature}
\label{sec:quarkmassBH}

The Unruh mechanism is believed to give an explanation for the dependence of the Unruh temperature ($T_{\mathrm{U}}$) on the acceleration ($a$), at which a particle are assumed to fly away from breaking of their strings taking place at the black-hole horizon; $T_{\mathrm{U}}=a/2 \pi$ \cite{tuchin}. The Hawking-Unruh temperature associated with the strange particle production which is obviously sensitive to the number of strange quarks in the hadrons, Fig. \ref{BH_Tc_Mq} (especially when assuming vanishing masses of the light quarks) plays a crucial role as its mass is nearly equivalent to the critical temperature \cite{1403.3541}. It was concluded that the chemical freezeout temperature is indirectly proportional to the quark masses \cite{1403.3541},
\bea
T_{\mathrm{U}} \simeq {\sigma \over \pi \sum_q\, w_q}, \label{Tq}
\eea
where $q$ runs over the quark flavors and  $w_q =m_q^2 + [\sigma^2/(4 m_q^2 + 2\pi \sigma)]^{1/2}$. In light of this, different freezeout temperatures can be estimated for different strange quarks, such as  $T_{\mathrm{U}}(00)=T_{\mathrm{U}}(000)$, $T_{\mathrm{U}}(0s)$, $T_{\mathrm{U}}(ss)=T_{\mathrm{U}}(sss)$, $T_{\mathrm{U}}(00s)$ and $T_{\mathrm{U}}(0ss)$ (see Table 1 in Ref. \cite{1403.3541}) at $m_s=0.1~$GeV and $\mu=0$. In determining these temperatures, it was assumed that the masses of the light quarks are vanishing. It has been noticed that the Hawking-Unruh temperature at $Q=0$ which is correspondent to the freezeout temperature  at $\mu=0$, i.e.  $T_{\mathrm{BH}}(M,Q=0)$, decreases with increasing the number of the strange quarks \cite{1403.3541}. The same observation is also confirmed in hadron resonance gas (HRG) model, Fig. \ref{BH_Tc_Mq}.

Furthermore, we have calculated the freezeout temperatures in dependence on different strange-quark contents from the HRG model at $\mu\neq0$ and confronted these with the results obtained from charged black-hole \cite{Tawfik:2015fda}, Fig. \ref{HawkingUnruhFreezeout}. We find that the proposed dependence of the freezeout temperatures on the strange-quark contents \cite{1403.3541} remains valid at $\mu\neq0$. This gives us a motivation to explain the strangeness production in high-energy collisions at finite baryon chemical potential.

\subsection{Freezeout temperatures from exact-string black hole}
\label{sec:tempExact}

As shall be discussed below, the exact-string black hole plays a crucial role in assuming that the thermal hadronization can be understood as Hawking-Unruh radiation. Dijkgraaf, Verlinde, and Verlinde \cite{Dijkgraaf:1991ba} have considered the string theory in two-dimensional black hole and studied the spectrum of the conformal field theory (CFT). They have used these in order to compute the scattering of strings off two-dimensional black hole and shown that the string propagator seems to exhibit Hawking radiation. Grumiller \cite{Grumiller:2005sq} introduced a construction for the action for the exact-string black hole allowing this type of black holes to be analogous to all orders of string coupling $(\alpha')$. Some thermodynamic properties of this black hole, such as, mass, temperature and entropy, have been determined. As introduced in previous sections, the extension to finite baryon chemical potential shall be implemented in the present work.

As explained, we originally assume that the thermal radiation that takes place from electrically-charged black hole is equivalent to the hadronization process at finite baryon chemical potential. That the earlier is indeed similar to Dijkgraaf-Verlinde-Verlinde black hole is an additional evidence for the correctness of our approach.

Furthermore, the Rindler hadronization (freezeout) process \cite{exact_string,Grumiller:2005sq} gives an explanation on how the Hawking temperature depends on the black-hole mass ($M$). The Unruh-hadronization temperature depends on $M$ \cite{0612151}. Castorina, Grumiller, and Iorio \cite{exact_string} discussed how the freezeout temperature in thermal hadronization can be related to the center-of-mass energy ($\sqrt{s_{\mathrm{NN}}}$),
\begin{itemize}
 \item at high $\sqrt{s_{\mathrm{NN}}}$, it is assumed that 
 \begin{equation}
 M=\gamma \sqrt{s_{\mathrm{NN}}},
 \end{equation}
 where $\gamma$ is a small factor, 
\item while at low $\sqrt{s_{\mathrm{NN}}}$
\begin{equation}
 M(\sqrt{s_{\mathrm{NN}}})=\Gamma\left(\sqrt{\sigma/s_{\mathrm{NN}}}\right) \sqrt{s_{\mathrm{NN}}}, \label{eq:MvsSqrts}
 \end{equation}
 where $\Gamma\left(\sqrt{\sigma/s_{\mathrm{NN}}}\right)$ was proposed in Ref. \cite{exact_string} but not specifying it. In Eq. (\ref{eq:Gammsqrts}), we suggest a parametrization for $\Gamma\left(\sqrt{\sigma/s_{\mathrm{NN}}}\right)$.
 \end{itemize}
This parametrization, Appendix \ref{AppendixA}, alternatively reads
\be
M(\sqrt{s_{\mathrm{NN}}})=1.19 \sqrt{s_{\mathrm{NN}}}^{\;0.957}.
\label{M_mu}
\ee
It is conjectured to be valid at all values of $\sqrt{s_{\mathrm{NN}}}$. Thus, the freezeout temperature can be related to finite $\mu$ as \cite{exact_string}
\be
T_{\mathrm{f}}(\mu)=\left.T_{\mathrm{f}}\right|_{\mu=0} \sqrt{1-\frac{2 \sqrt{2 \pi \sigma}}{M(\mu)}}.
\label{T_mu_finite}
\ee
This expression is very similar to Eq. (4.6) in Ref. \cite{Grumiller:2005sq}, but here it is conjectured that $M$ is a function $\mu$, see Eq. (\ref{eq:bestfittings}). It is assumed that (see, for example \cite{exact_string})
\be
\left.T_{\mathrm{f}}\right|_{\mu=0}=\sqrt{\frac{\sigma}{2\pi \omega_q}}. \label{T_mu_0}
\ee
where $\omega_q$ the quark's effective mass, $\omega_q=(m_q^2 + k_q^2)^{1/2}$, with $m_q$ being the bare mass of the produced quark and $k_q$ being the quark's momentum inside the hadronic state $\bar{q}_1 q_1$ and $\bar{q}_2 q_2$. The latter describes a pair production mechanism or string breaking process. It was found that at vanishing $\omega_q$, $\left.T_{\mathrm{f}}\right|_{\mu=0}\simeq 170~$MeV \cite{0704.1426}.

From the statistical thermal fit of various particle ratios measured in different high-energy collision experiments, many authors have proposed  relations between the chemical potential ($\mu$) and the nucleon-nucleon center-of-mass energy ($\sqrt{s_{\mathrm{NN}}}$), see for instance Ref. \cite{Tawfik:2014eba} for recent review. In Eq. (\ref{M_mu}), it is straightforward to substitute $\sqrt{s_{\mathrm{NN}}}$ with $\mu$, which was proposed by different authors, see for instance Ref. \cite{Tawfik:2014eba}. This leads to an expression for the black-hole mass ($M$) as a function of the baryon chemical potential ($\mu$)
\be
M(\mu)=1.19 \left(\frac{a-\mu}{b \mu}\right)^{0.957}, \label{eq:bestfittings}
\ee
where $\mu$ is given in units of GeV and $a=1.245\pm0.094$ GeV and $b=0.264\pm0.028$ GeV$^{-1}$ are constants \cite{Tawfik:2013bza,Tawfik:2014dha}.

\subsection{Particle ratios at freezeout temperatures}
\label{sec:Ratios}

In the confined phase, different hadrons and their resonances treated as an ideal, noninteracting gas~\cite{Karsch:2003vd,Karsch:2003zq,Redlich:2004gp,Tawfik:2004sw,Tawfik:2004vv} are conjectured to come up with contributions to the {\it total} thermodynamic pressure, for instance. The possible correlations and strong interaction seem to be taken into consideration in such a grand-canonical ensemble which is composed of recent compilation by the Particle Data Group (PDG) with masses $\le 2~$GeV \cite{Tawfik:2004sw}
\begin{eqnarray}
Z(T, \mu, V) &=& \mathtt{Tr} \left[ \exp^{\frac{\mu\, N-H}{T}}\right],
\end{eqnarray}
where $H$ is the Hamiltonian of the system, which is given as a summation of the kinetic energies of the relativistic Fermi and Bose particles, counting for the degrees of freedom of the confined hadrons and thus implicitly including various types of interactions, at least the ones responsible for the formation of further resonances and so on. This model gives an excellent description for the particle production in the heavy-ion collisions \cite{Tawfik:2014eba}. With these assumptions, the dynamics of the partition function can be calculated as sum over  {\it single-particle partition} functions ($Z_i^1$) of $i$-th hadron or resonance
\begin{eqnarray} 
\ln Z(T, \mu_i ,V) &=& 
\pm \sum_i \frac{V g_i}{2\pi^2}\int_0^{\infty} k^2 dk \ln\left\{1 \pm \exp[(\mu_i -\varepsilon_i)/T]\right\}, \label{eq:lnz1}
\end{eqnarray}
where $\varepsilon_i(k)=(k^2+ m_i^2)^{1/2}$ is the $i-$th particle dispersion relation, $g_i$ is
spin-isospin degeneracy factor and $\pm$ stands for fermions and bosons, respectively.

The particle number can be deduced as \cite{Tawfik:2013bza,Wheaton:2009}
\begin{equation}
n(T,\mu)= \sum_i \frac{g_i}{2 \pi^2} T m_i^2 K_2\left(\frac{m_i}{T}\right) \gamma_s^{(n_s)_i} \exp\left(\frac{-\mu_i}{T}\right), \label{eq:n1}
\end{equation}
where $\mu_i=(n_b)_i\;\mu_b + (n_s)_i\;\mu_s$, $(n_b)_i$ and $(n_s)_i$ being the $i$-th particle's baryon and strange quantum number, respectively. $\mu_b$ is the baryon chemical potential and $\mu_s$ is the strangeness one. The influence of strange quark on  the QCD phase-diagram and the chemical freezeout has been analysed in Ref. \cite{Tawfik:2004vv}. It was found that the freezeout temperature decreases with the increase in the number of the strange quarks. This observation is very compatible with the QCD results, which are confirmed in the HRG freezeout temperatures \cite{Tawfik:2014eba}.  Fig. \ref{BH_Tc_Mq} depicts these results. The same conclusion can be drawn from the black-hole radiation, see Fig. \ref{HawkingUnruhFreezeout} and section \ref{sec:quarkmassBH}.

In the present work, calculations of three different particle ratios under the assumption that the thermal hadronization can be interpreted as Hawking-Unruh radiation at vanishing and finite baryon chemical potential shall be presented. In other words, we extend the calculation of $\mathrm{K}^+/\pi^+$ which was introduced in Ref. \cite{1403.3541} to finite chemical potential. Furthermore, we introduce calculations for the ratio of the strange mesons $\phi$ to $\mathrm{K}^-$ and the ratio of strange anti-baryon $\bar{\Lambda}$ to non-strange meson $\pi^-$ at vanishing and finite baryon chemical potential.

From HRG, Eq. (\ref{eq:n1}), the final state particle number is composed of contribution from {\it stable} hadrons and additional ones from heavier resonances which decay into the $i$-th particle of interest,
\bea
\left.\langle n_{(i)}\rangle\right|_{\mathrm{final}} &=& \left.\langle n_{(i)}\rangle\right|_{\mathrm{stable}} + \sum_{j=1}\; b_{j\rightarrow i}\; \langle n_{(j)}\rangle, \label{eq:nBr}
\eea
where $b_{j\rightarrow i}$ is the branching ratio for $j$-th resonance decaying into the $i$-th particle. The particle ratio $N_{\mathrm{K}^+}/N_{\pi^+}$ can be given as
\be
\frac{N_{\mathrm{K}^+}}{N_{\pi^+}}=\sum_i \frac{g_{\mathrm{K}^+}\; m_{\mathrm{K}^+}^2}{g_{\pi^+} \; m_{\pi^+}^2} \gamma_s  \frac{K_2\left(\frac{m_{\mathrm{K}^+}}{T}\right)}{K_2\left(\frac{m_{\pi^+}}{T}\right)} \exp\left(\frac{-n_s \mu_s}{T}\right), \label{eq:kpi}
\ee
with $N$ measures the mean multiplicity. In calculating Eq. (\ref{eq:kpi}), the additional contributions expressed in Eq. (\ref{eq:nBr}) are taken into consideration. 


From Hawking-Unruh radiation, the freezeout temperature corresponding to the kaons was defined as $T_{\mathrm{K}}=T(0s)$ \cite{1403.3541}. For pion, $T_\pi=T(00)$. Both values are distinguishable, especially at low $\mu$. It is obvious that $T_\pi$ is nothing but the freezeout temperature at vanishing chemical potential, Eq. (\ref{T_mu_0}), i.e.  no strange quark.  From the statistical treatment of the Hawking-Unruh radiation, the strange quark occupation parameter ($\gamma_s$) can be omitted,
\be
\frac{N_{\mathrm{K}^+}}{N_{\pi^+}}=\sum_i \frac{g_{\mathrm{K}^+}\; m_{\mathrm{K}^+}^2}{g_{\pi^+} \; m_{\pi^+}^2} \frac{T_{\mathrm{K}^+}}{T_{\pi^+}}  \frac{K_2\left(\frac{m_{\mathrm{K}^+}}{T_{\mathrm{K}^+}}\right)}{K_2\left(\frac{m_{\pi^+}}{T_{\pi^+}}\right)}\exp\left(\frac{-n_s \mu_s}{T_{\pi^+}}\right).
\label{Koverpi}
\ee
At $\sigma=0.2$ GeV$^2$, $m_s=0.1$ GeV, $T_\pi=178$ MeV, $T_K=167$ MeV (see Table 1 in Ref. \cite{1403.3541}) and at finite $\mu$ ($\mu \neq 0$), this ratio is depicted in Fig. \ref{BH_Kp_pip} as a function of $\sqrt{s_{\mathrm{NN}}}$.

It is straightforward to give similar expressions for $\phi/{\mathrm{K}}^-$ and $\bar{\Lambda}/\pi^-$,
\bea
\frac{N_\phi}{N_{\mathrm{K}^-}}&=& \sum_i \frac{g_\phi\; m_\phi^2}{g_{\mathrm{K}^-} \; m_{\mathrm{K}^-}^2} \frac{T_\phi}{T_{\mathrm{K}^-}}  \frac{K_2\left(\frac{m_\phi}{T_\phi}\right)}{K_2\left(\frac{m_{\mathrm{K}^-}}{T_{\mathrm{K}^-}}\right)}\exp\left(\frac{-n_s \mu_s}{T_{\mathrm{K}^-}}\right), \label{Phioverk} \\
\frac{N_{\bar{\Lambda}}}{N_{\pi^-}}&=& \sum_i \frac{g_{\bar{\Lambda}}\; m_{\bar{\Lambda}}^2}{g_{\pi^-} \; m_{\pi^-}^2} \frac{T_{\bar{\Lambda}}}{T_{\pi^-}}  \frac{K_2\left(\frac{m_{\bar{\Lambda}}}{T_{\bar{\Lambda}}}\right)}{K_2\left(\frac{m_{\pi^-}}{T_{\pi^-}}\right)}\exp\left(\frac{-n_s \mu_s+n_b \mu_b}{T_{\pi^-}}\right). \label{aLaoverpi}
\eea
Left-hand panel of Fig. \ref{BH_Phi_km} shows the results on $N_{\phi}/N_{\mathrm{K}^-}$ at $T_\phi=T(ss)=157$ MeV, while right-hand panel presents $N_{\bar{\Lambda}}/N_{\pi^-}$ at $T_{\bar{\Lambda}}=T(00s)=171$ MeV.

For a realistic comparison of the particle ratios, heavy resonances decaying into the particle of interest should take into account. In the black-hole approach, the abundances of some heavy resonances have been determined in Ref. \cite{newRef}. The authors concluded that, the suppression of strangeness production and the perturbation from equilibrium are likely due to the thermal behavior of the color confinement event horizon and due to the dependence of the resulting freezeout temperatures on the strangeness quantum number of the produced particles of interest  \cite{1403.3541}. When comparing their results from string tension and bare strange quark masses, a good agreement with the measured hadronic abundances was obtained.  As expressed in Eq. (\ref{eq:nBr}) considerable contributions should be added to the {\it primary} ones given in  Eqs. (\ref{Koverpi})-(\ref{aLaoverpi}).

In our calculations, we make use of the freezeout temperatures proposed in Ref. \cite{1403.3541} at vanishing chemical potential. In other words, $T_{\mathrm{\pi}^\pm}$, $T_{\mathrm{K}^\pm}$, etc. vary with increasing $\mu$. This is a main assumption in the present work. At $\mu\neq0$, the freezeout temperatures for pion, kaon, etc. can be calculated from Eqs. (\ref{T-Q}) and (\ref{T_mu_finite}). The latter inversely relates the freezeout temperature of given particle species to the black-hole mass as a function of $\mu$, Eq. (\ref{eq:bestfittings}).

\section{Results and Discussion}
\label{sec:res}

\begin{figure}[h]
\includegraphics[width=8.cm,angle=-90]{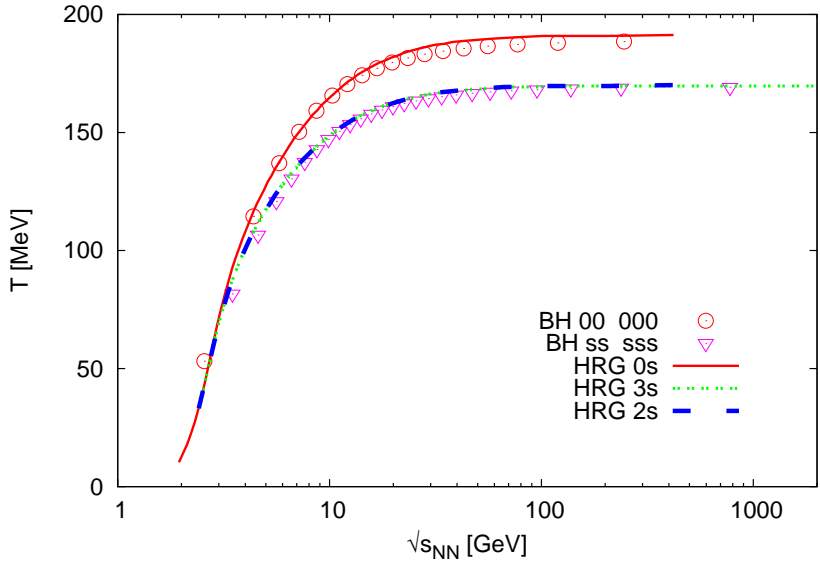}
\caption{(Color online)  The freezeout temperatures are given as functions of $\sqrt{s_{\mathrm{NN}}}$. Symbols represent results for the Hawking-Unruh formulations, Eq. (\ref{T-Q}). The curves give the HRG calculations, Eq. (\ref{eq:lnz1}), at $s/T^3=7$. 
\label{BH_Tc_Mq}
}
\end{figure}

In Fig.  \ref{BH_Tc_Mq}, the freezeout temperatures are calculated at varying $\sqrt{s_{\mathrm{NN}}}$, i.e. at $\mu\neq 0$. Results from the Hawking-Unruh thermal radiation at $\mu\neq 0$ (Symbols), Eq. (\ref{T-Q}), are confronted to the HRG calculations, Eq. (\ref{eq:lnz1}), which are estimated at the freezeout condition $s/T^3=7$ \cite{Tawfik:2005qn,Tawfik:2004ss}. There is an excellent agreement between black holes and HRG with no strange quarks. Also, both calculations with two and three strange quarks are very compatible with each others and among themselves. The third observation to be highlighted is the strong dependence of the freezeout temperatures  on the strange quark contents of the produced hadrons.

\begin{figure}[h]
\includegraphics[width=10.cm]{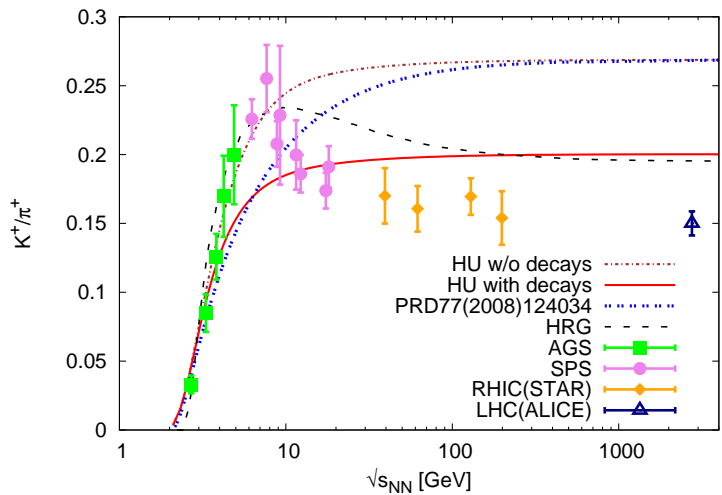}
\caption{(Color online) The dependence of $\mathrm{K}^+/\pi^+$ ratio on $\sqrt{s_{\mathrm{NN}}}$, which is inversely proportional to $\mu$ or equivalently to $Q$, is depicted. The calculations from Eq. (\ref{T-Q}) \cite{Tawfik:2015fda} are given as solid curve dot-dashed curves without and with resonance decays, respectively, and that from Eq. (\ref{T_mu_finite}) \cite{exact_string} are depicted as double-dotted curve. The dashed curve presents the calculations from the HRG model. The experimental results from AGS \cite{L.Ahle:2000}, SPS \cite{I.G.Bearden:2002,S.Afanasiev:2002,M.Gazdzicki:2004}, RHIC (STAR) \cite{J.Adams:2005} and LHC (ALICE) \cite{B.Abelev:2012} are shown as solid squares, solid circles, solid diamonds and open triangles, respectively. }
\label{BH_Kp_pip}
\end{figure}

Fig. \ref{BH_Kp_pip} shows estimations for $\mathrm{K}^+/\pi^+$ as a function of $\sqrt{s_{\mathrm{NN}}}$, which is inversely proportional to $\mu$ (electric charge of the black holes) \cite{Tawfik:2014eba,Tawfik:2013bza,Tawfik:2014dha}. The solid and dot-dashed curves represent our calculations from Eq. (\ref{Koverpi}) with and without resonance decays, respectively. The calculations based on exact-string black hole, section \ref{sec:tempExact}, are depicted as double-dotted curve. The dashed curve shows the calculations from HRG (with resonance decays) based on constant entropy density normalised to $T^3$; $s/T^3=7$ \cite{Tawfik:2005qn,Tawfik:2004ss}. Alternating gradient synchrotron (AGS) \cite{L.Ahle:2000}, SPS \cite{I.G.Bearden:2002,S.Afanasiev:2002,M.Gazdzicki:2004}, relativistic heavy-ion collider (RHIC) (STAR experiment) \cite{J.Adams:2005} and large hadron collider (LHC) (ALICE experiment) \cite{B.Abelev:2012} measurements are given as solid squares, solid circles, solid diamonds and open triangle, respectively.

There is an excellent agreement between the $\mathrm{K}^+/\pi^+$ calculations at varying $\sqrt{s_{\mathrm{NN}}}$, which are based on black-hole radiation given by the dot-dashed curve. These agree well with the thermal HRG estimations at the universal freezeout condition $s/T^3=7$, (dashed curve) and with the phenomenologically deduced ratios (symbols), especially at low energies. At RHIC and LHC energies, the calculations based on exact-string black hole greatly overestimate the measurements. They are in good agreement with the present calculations when the resonance decays are not taken into consideration.
On the other hand, at RHIC and LHC energies, the present calculations with the resonance decays excellently agree with the HRG results. Indeed both still overestimate the measurements. To summarize we highlight that, at high energies, the present calculations with the resonance decays and the HRG results are almost identical. They are closer to the measurements than the exact-string and our calculations without resonances decays. Thus, we might conclude that the resonance decays seem to play a crucial role in this energy scale, while they might be ignored at lower energies. Additional to them, other ingredients should be taken into account.

As introduced in section \ref{sec:Ratios}, HRG assumes ideal, point-like constituents (hadron resonances). The HRG calculations for $\mathrm{K}^+/\pi^+$ ratio are determined as a thermal averaging over different decay channels in addition to the stable particles themselves, while in the black-hole calculations, Eq. (\ref{Koverpi}), no such averaging is applied. Also, it is assumed that the light and strange quark occupation parameters, $\gamma_q$ and $\gamma_s$, respectively, are unity, i.e. equilibrium processes. They might play a role to bring the calculations closer to the measurements. Nevertheless, it is not the scope of this paper to discuss the reproductivity of the freezeout parameters, which phenomenologically are deduced from the statistical fit of the particle ratios from the HRG model assuming universal freezeout conditions, such as $s/T^3=7$ \cite{Tawfik:2014eba}. 

\begin{figure}[h]
\includegraphics[width=10.cm]{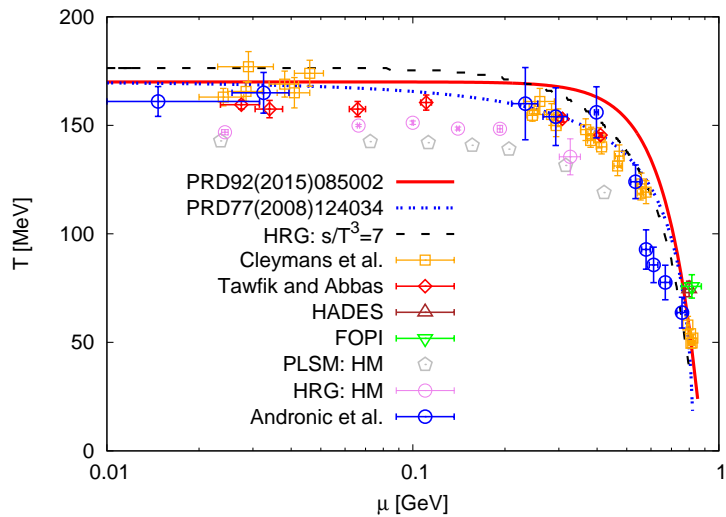}
\caption{(Color online) The freezeout parameters $T$ and $\mu$ (solid curve) \cite{Tawfik:2015fda} determined from the analogy of Hawking-Unruh radiation to the QCD particle production  are depicted. The solid curve presents the present calculations (without resonances decay). The HRG calculations are illustrated by dashed curve. The double-dotted curve shows the results from the proposed Eqs. (\ref{eq:MvsSqrts}) and (\ref{eq:bestfittings}) for exact-string black hole. The symbols refer to the experimentally deduced freezeout parameters from particle ratios: Cleymans {\it et al.} (open square) \cite{clmns}, Tawfik {\it et al.} (open diamond) \cite{Tawfik:2013bza,Tawfik:2014dha}, HADES (open triangle) \cite{hds}, and FOPI \cite{fopi} and higher-order moments: SU(3) Polyakov linear-ó model (PLSM)  and HRG \cite{fohm}. We need additional curves depicting our results with resonances decays!}
\label{HawkingUnruhFreezeout}
\end{figure}

At high energy, the calculations from the exact-string black holes, Eq. (\ref{T_mu_finite}) \cite{exact_string}, obviously overestimate both HRG and experimental results. Bu they are in good agreement with our calculations when the resonance decays are not included, Eq. (\ref{Koverpi}) with Eq. (\ref{T-Q}) \cite{Tawfik:2015fda}.  At $\mu=0$, the freezeout temperatures can be determined from Eq. (\ref{Tq}) \cite{1403.3541}. Although, the two types of calculations are almost identical, within the $\mu$-range, where the freezeout temperature rapidly decreases, they become distinguishable. This can be interpreted from the dependence of the freezeout temperature on the black-hole mass, Eq. (\ref{T_mu_finite}). In charged black-hole, the mass depends on the black-hole electric charge, Eq. (\ref{eq:1stThrm}). On the other hand, in the exact-string black holes, the mass depends on $\mu$ which stems from the proposed function for the relation between the mass and $\sqrt{s_{\mathrm{NN}}}$, Eq. (\ref{eq:MvsSqrts}). Another reason for the distinguishable results comes from the differences in the temperature dependence on black-hole charge and on the chemical potential, Fig. \ref{HawkingUnruhFreezeout}. We notice that the further increase in $\mu$ brings both calculations closer to each other.
 
We observe that at high energy, the black-hole calculations for $\mathrm{K}^+/\pi^+$ (without decays) keep the largest value measured at top SPS energies (the value reported in Ref. \cite{1403.3541}) unchanged, while the measurements in different experiments seem to decrease with increasing $\sqrt{s_{\mathrm{NN}}}$. At very high energy, the value of  $\mathrm{K}^+/\pi^+$ seems no longer dependent on $\sqrt{s_{\mathrm{NN}}}$, but both HRG and black-hole calculations overestimate the measurements. At low energy, both calculations and measurements rapidly decrease. Thus, we conclude that, although the fair qualitative agreement, neither strangeness suppression nor enhancement is found. For their thermal characterization, the heavy resonances come up with considerable ingredients.

\begin{figure}[h]
\includegraphics[width=8.cm]{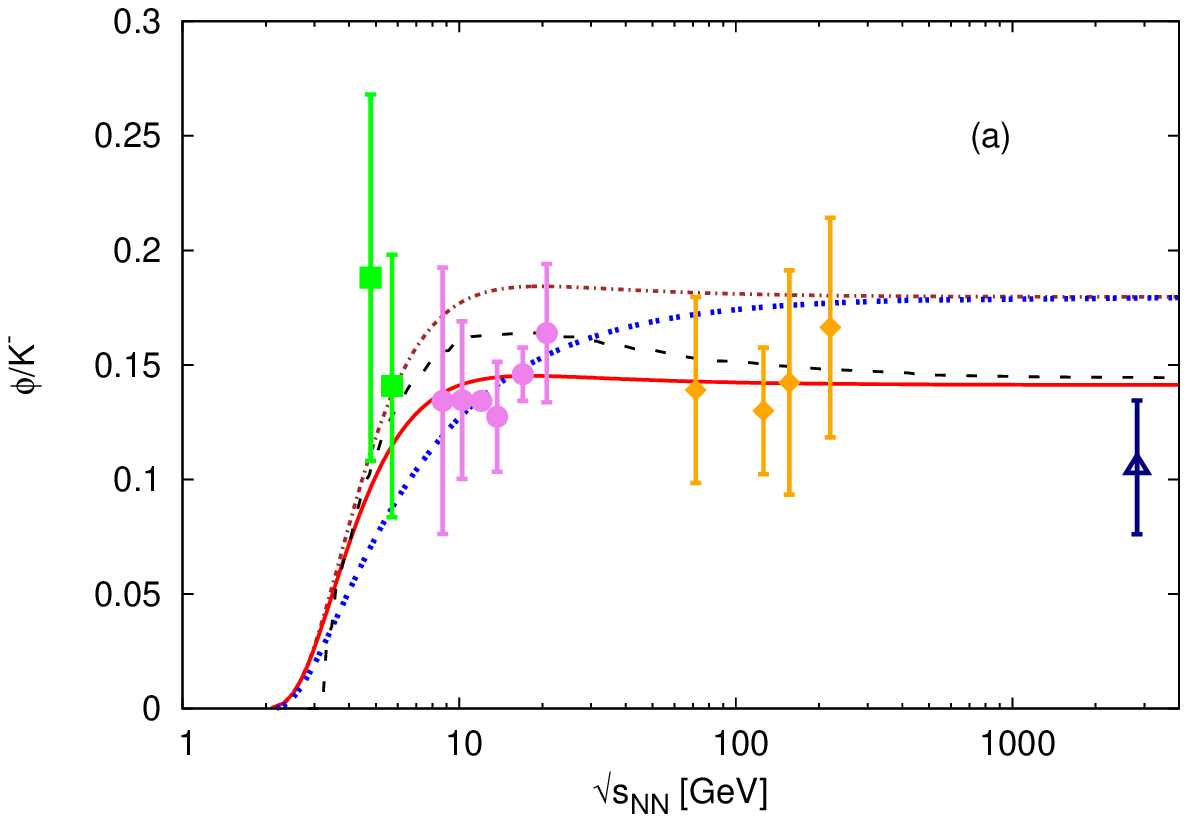}
\includegraphics[width=8.cm]{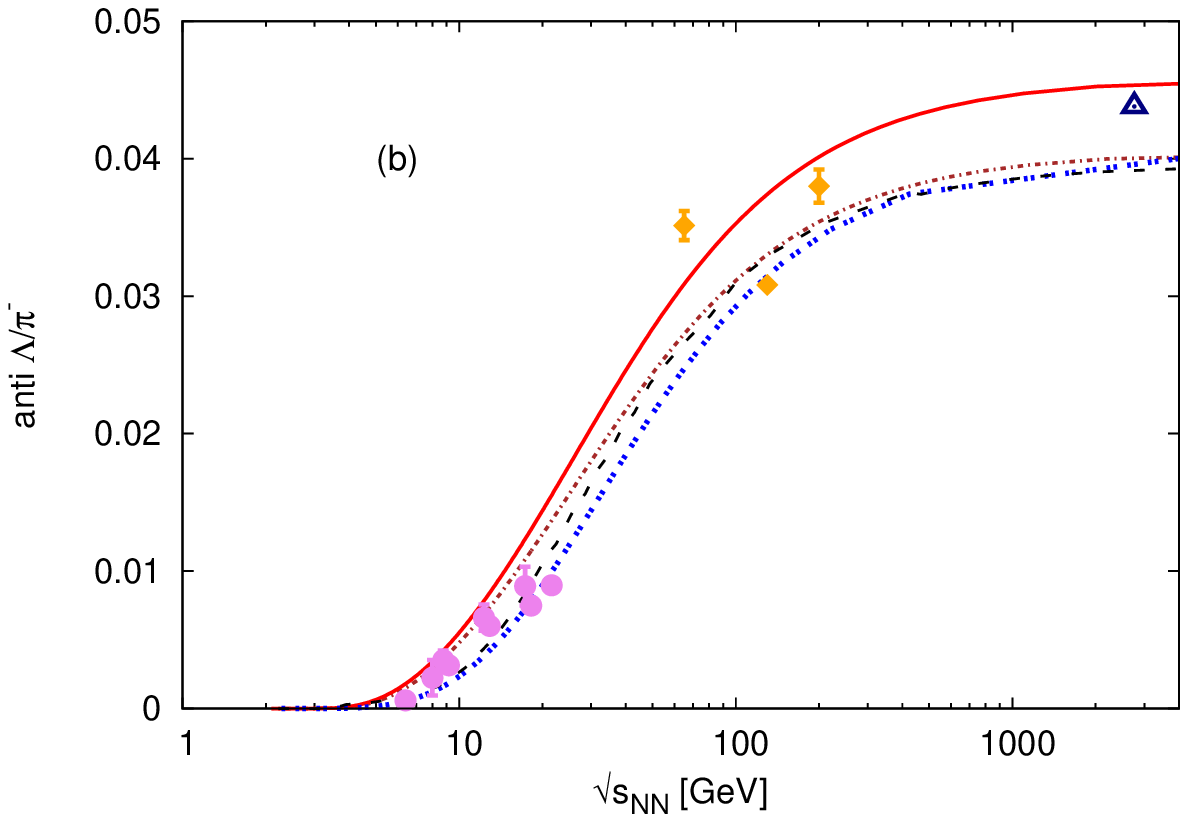}
\caption{(Color online) The same as in Fig. \ref{BH_Kp_pip} but for $\phi/\mathrm{K}^-$ (left-hand panel) and ${\bar{\Lambda}}/\pi^-$  (right-hand panel).}
\label{BH_Phi_km}
\end{figure}

Fig. \ref{BH_Phi_km} shows the $\phi/\mathrm{K}^-$ (left-hand panel) and ${\bar{\Lambda}}/\pi^-$  (right-hand panel) as functions of $\sqrt{s_{\mathrm{NN}}}$. The solid line shows the present work for charged black-hole results as calculated from Eq. (\ref{aLaoverpi}) when resonance decays are considered. The resonances decays are taken into account. The dot-dashed curve presents the calculated ratio by using exact-string black hole relations \cite{exact_string}. The dashed curve stands for the HRG results (with resonances decays) while the experimental data [SPS \cite{T.Anticic:2004,F.Antinori:2004}, RHIC (STAR) \cite{J.Adams:2005} and LHC (ALICE) \cite{B.Abelev:2013}] are also given as symbols.

The absence of strangeness suppression especially at high energies as reported in Fig. \ref{BH_Kp_pip} is also observed in Fig. \ref{BH_Phi_km}, but the possible enhancement at top SPS energies is not confirmed by the black-hole calculations. At lower energies, the value of $\phi/\mathrm{K}^-$, Eq. (\ref{Phioverk}), rapidly decreases while the measurements in different experiments - not shown here due to radical difference in their rapidity cuts - seem to signal a rapid increase \cite{Tawfik:2014eba}.

At high energies, there is a good agreement between the HRG and the measurements. The black-hole  estimations agree well, especially our present calculations with resonances decays (solid curves). As observed in Fig. \ref{BH_Kp_pip}, the calculations based on black-hole radiation when resonances decays are not considered seems to overestimate both HRG and the experiments. Here, the overestimation is obviously smaller than in Fig. \ref{BH_Kp_pip}. For ${\bar{\Lambda}}/\pi^-$, the agreement is fairly good. All approaches seem to agree with each others.

\begin{figure}[h]
\includegraphics[width=10.cm]{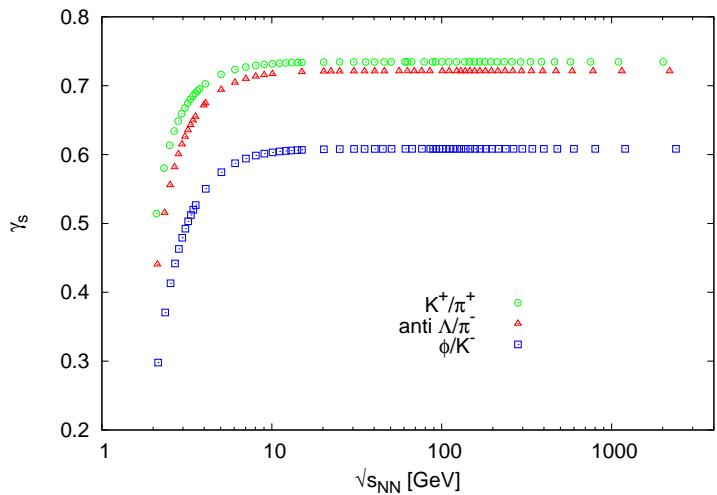}
\caption{(Color online) The dependence of the strangeness suppression factor $\gamma_s$ on $\sqrt{s_{\mathrm{NN}}}$ for the three particle ratios, $\mathrm{K}^+/\pi^+$ (circle symbol), $\bar{\Lambda}/\pi^-$ (triangle symbol) and $\phi/\mathrm{K}^-$ (square symbol). }
\label{gammas}
\end{figure}

We observe that  the strangeness production is obviously enhanced with $\sqrt{s_{\mathrm{NN}}}$, Fig. \ref{gammas}.  It is obvious that $\gamma_s$ rapidly increases with $\sqrt{s_{\mathrm{NN}}}$, especially at low energies such as AGS and low SPS. At high energies, $\gamma_s$ saturates. The energy at which the energy-dependent $\gamma_s$ converts to an energy-independent function coincides with top SPS energy. This energy region corresponds to the onset of deconfinement phase-transition and the transition from baryon- to meson-dominated HRG \cite{Tawfik:2014eba}. This result has led to suspect that QGP might have been produced at this low energy \cite{QGPearly}. Such a conclusion which is based on macroscopic observations, ignores the fact that the hadronic and partonic degrees-of-freedom should be distinguishably accessible and their impacts on the final-state particle production should be fully-microscopically characterized. Furthermore, we notice that $\gamma_s$ decreases with increasing the number of strange quarks.

Fig. \ref{HawkingUnruhFreezeout} illustrates the freezeout plane ($T$ vs. $\mu$) \cite{Tawfik:2015fda} determined from the analogy between Hawking-Unruh radiation and the particle production in high-energy experiments (hadronization). The solid curve depicts the present calculations (without resonances decay). The results deduced from Eqs. (\ref{eq:MvsSqrts}) and (\ref{eq:bestfittings})  for exact-string black hole which are calculated from Eq. (\ref{T_mu_finite}) are compared with the results from the HRG model at the freezeout condition $s/T^3=7$ (dashed curve) and the phenomenologically deduced parameters (symbols) from the particle ratios: Cleymans {\it et al.} \cite{clmns}, Tawfik {\it et al.}  \cite{Tawfik:2013bza,Tawfik:2014dha}, HADES \cite{hds}, and FOPI \cite{fopi} and the higher-order moments:  SU(3) Polyakov linear-sigma model (PLSM) and HRG \cite{fohm}.

We find an excellent agreement between HRG \cite{Tawfik:2014eba}, electrically-charged black hole \cite{Tawfik:2015fda} and exact-string black hole \cite{exact_string} results on the freezeout temperature ($T$) as functions of $\sqrt{s_{\mathrm{NN}}}$. As discussed in section \ref{sec:tempExact}, the dependence of $T$ on $\sqrt{s_{\mathrm{NN}}}$ is sensitive to the proposed proportionality function $\Gamma$, which is approximately determined in Appendix \ref{AppendixA}. The excellent agreement between the HRG and the charged black-hole results means that our estimation for $\Gamma(\sqrt{\sigma/s_{\mathrm{NN}}})$ correctly describes the proportionality of the freezeout temperature to $\sqrt{s_{\mathrm{NN}}}$. The small difference between these calculations and the ones reported in \cite{Tawfik:2015fda}, especially at low energy, has been discussed earlier.

\section{Conclusions}
\label{sec:cncl}

The possible enhancement of the strange particle production in $A-A$ relative to $p-p$ collisions was proposed as a probe for dynamics of heavy-ion collisions and their degrees of freedom and thus for one of the earliest signatures for the QGP formation \cite{ReF1,ReF2}. It is conjectured that the deconfined state of free quarks and gluons rapidly approaches the chemical equilibrium and thus can be described as a grand-canonical ensemble. This implies a considerable increase in the strange-quark phase-space occupancy parameter ($\gamma_s$) relative to the one observed in $p-p$ collisions. The large production yields observed in $A-A$ collisions at SPS \cite{ReF15,ReF16} and RHIC \cite{ReF17} energies, especially the multi-strange baryons, can not be interpreted as stemming from a hadronic phase due to lack of fast equilibration.

The strange and multi-strange particles have been measured in various $p-p$ and $A-A$ experiments \cite{Alice2013}. It was found that the $A-A$ strange particle yields normalized to the corresponding yields from the $p-p$ collisions are enhanced with increasing the collision centrality.  Furthermore, such an enhancement is obviously reduced with increasing the strange quark contents of the particles of interest. This is consistent with the enhancement in the $\bar{s}s$ pair-production in hot and dense QCD medium. It has been found that the enhancement generally decreases with increasing $\sqrt{s_{\mathrm{NN}}}$.

An enhancement instead of a suppression was measured at low energies \cite{1403.6311}. This finding has been interpreted due to the assumption of chemical nonequilibrium of the strange particles, i.e.  $\gamma_s \simeq 1.2-1.6$. As introduced in Ref. \cite{0511092}, $\gamma_s$ seems to increase from $0.45$ in $p-p$ collisions to $0.8$ in $Pb-Pb$ collisions at the same energy ($17.2~$GeV), while the dependence of $\gamma_s$ on $\sqrt{s_{\mathrm{NN}}}$ is found to be much moderate. It varies from $0.65$ to $0.84$ over the energy rage $4.7$ to $12.7$ GeV.

In the present work, we propose that the ratios of various particle spices produced from heavy-ion collisions at various energies can be utilized in estimating  the freezeout temperatures and chemical potentials. Furthermore, they are correspondent to the particle production from the black-hole radiation (charged and exact-string black hole). We have studied the influence of the number of strange quarks in the particle of interest on the variation of the freezeout temperature with varying baryon chemical potential. The chemical freezeout parameters were related to the Hawking-Unruh radiation from electrically-charged black holes. We also compared the resulting freezeout temperature from exact-string black holes by using a proposed proportionality of the black-hole mass to the center-of-mass energy (or $\mu$) with the freezeout parameters deduced from the ratios of various particle species from heavy-ion collision experiments. An excellent agreement is found. We extend the inverse dependence of the freezeout temperature on the number of strange particles at finite chemical potential and found a convincing agreement, as well.

Furthermore, we have estimated three different particle ratios by using two types of black holes (charged and exact string) at varying center-of-mass energies or finite chemical potentials and compared the results with the HRG model and different measurements, Figs. \ref{BH_Kp_pip} and \ref{BH_Phi_km}. For $\mathrm{K}^+/\pi^+$, an excellent agreement between their measurements and calculations at varying $\sqrt{s_{\mathrm{NN}}}$ is found. At RHIC and LHC energies, the exact-string black holes greatly  overestimate the experimental results. They agree well with our calculations without resonances decays. The calculations with resonance decays agree well with HRG. Both are closer the measurements than exact-string and even our calculations without resonances decays. This leads to the conclusion that the resonance decays are essential at high energies (low $\mu$), while at lower energies, they seem not playing the same crucial role.

The Hawking-Unruh thermal radiation is conjectured as the counterpart mechanism to the hadronization process in high-energy collisions. This mechanism assumes that a small number of quarks and gluons is connected even in the elementary collisions and the particle production takes place from the sequential breaking of quark-antiquark strings as tunnelling processes through the event horizon of the color confinement. We conclude that the temperature and chemical potential play a relevant role in the strangeness suppression.

\appendix
\section{Proportionality of black-hole mass to $\sqrt{s_{\mathrm{NN}}}$}
\label{AppendixA}

Through the relation $M = \gamma \sqrt{s_{\mathrm{NN}}}$, the black-hole mass ($M$) is to be related to nucleon-nucleon center-of-mass energy ($\sqrt{s_{\mathrm{NN}}}$), at high energy, where $\gamma$ is a proportionality function \cite{exact_string} which becomes valid only at high $\sqrt{s_{\mathrm{NN}}}$. At low energy, Ref. \cite{exact_string} suggested that the dependence of $M$ on $\sqrt{s_{\mathrm{NN}}}$ reads
\bea
M=\Gamma\left(\sqrt{\sigma/s_{\mathrm{NN}}}\right) \sqrt{s_{\mathrm{NN}}}. \label{eq:MGammSqrtS}
\eea
We propose that the proportionality function is given as
\bea
\Gamma\left(\sqrt{\sigma/s_{\mathrm{NN}}}\right)= \sqrt{\frac{2\pi\sigma}{s_{\mathrm{NN}}}} + \gamma \tanh\left(\frac{0.5}{\sqrt{2\pi\sigma/\sqrt{s_{\mathrm{NN}}}}}\right). \label{eq:Gammsqrts}
\eea
By fitting the results which are obtained from the proportionality function, Eq. (\ref{eq:Gammsqrts}) then
\bea
M(\sqrt{s_{\mathrm{NN}}})=1.19 \sqrt{s_{\mathrm{NN}}}^{\;0.957}.
\eea

\begin{figure}[h]
\includegraphics[width=8.cm]{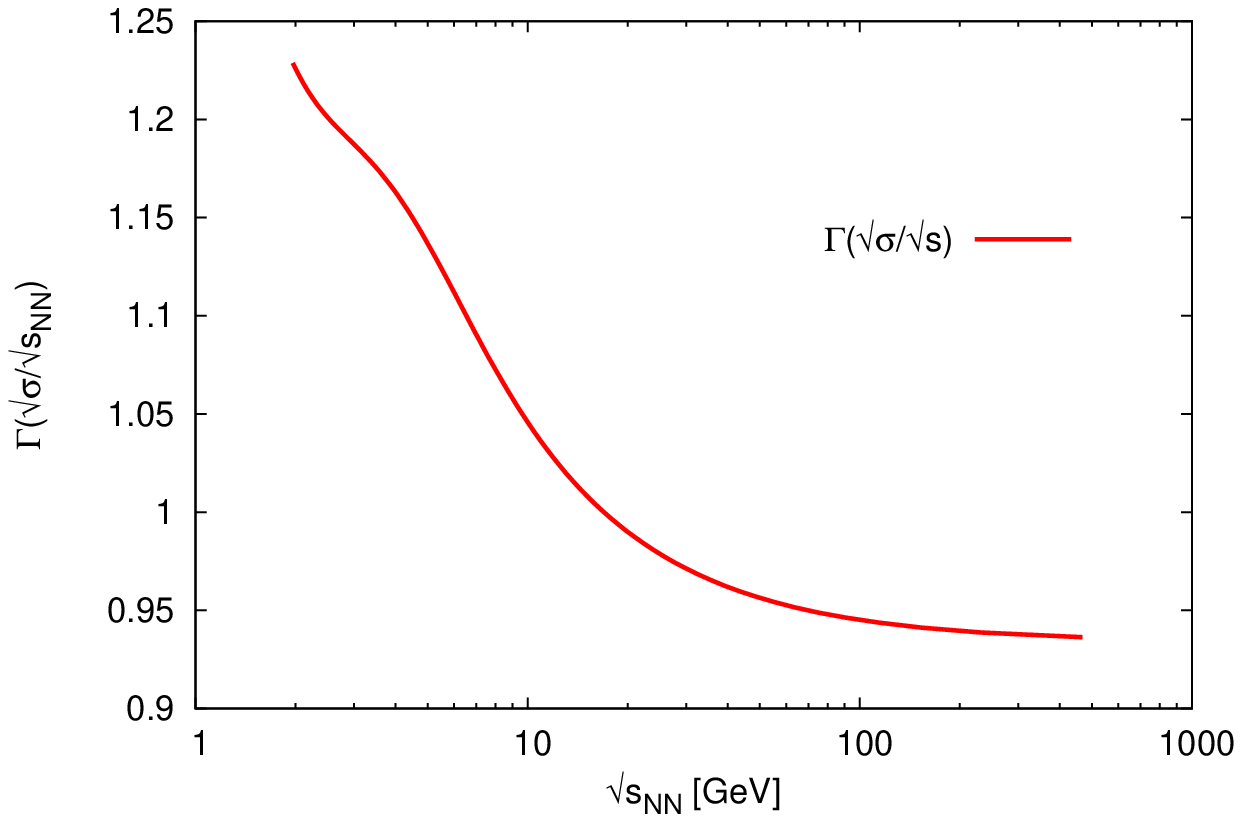}
\includegraphics[width=8.cm]{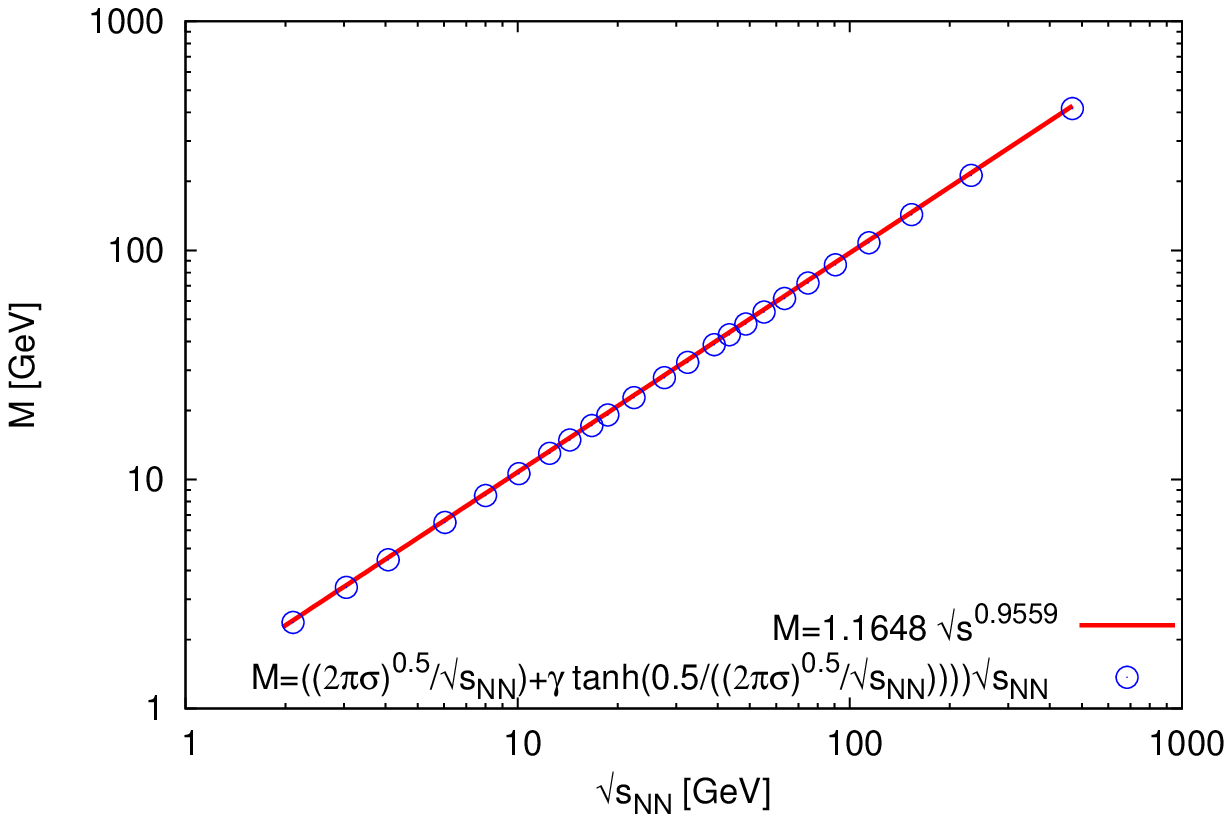}
\caption{Left-hand panel: the proportionality function relating black-hole mass to $\sqrt{s_{\mathrm{NN}}}$ in Eq. (\ref{eq:MvsSqrts}) is calculated as a function of $\sqrt{s_{\mathrm{NN}}}$ in Eq. (\ref{eq:Gammsqrts}). Right-hand panel depicts Eq. (\ref{eq:MvsSqrts}) with $\Gamma(\sqrt{s_{\mathrm{NN}}})$ given in the left-hand panel. The solid curve present the best fitting, Eq. (\ref{eq:bestfittings}).  }
\label{Mass_with_sqrtS}
\end{figure}

Fig. \ref{Mass_with_sqrtS} shows the relation between the proposed and approximated proportionality function in dependence on $\sqrt{s_{\mathrm{NN}}}$. We find an excellent agreement between the two expressions due to the small value of chi-squared ($\chi^2$). It is obvious that the fit describes well the results. Also, the correctness of this proposed function can be confirmed, even graphically in Fig. \ref{Mass_with_sqrtS}.

\end{document}